\documentclass[12pt,preprint]{aastex}

\def\ifundefined#1{\expandafter\ifx\csname#1\endcsname\relax}

\newif\ifpdf
\ifx\pdfoutput\undefined
   \pdffalse      
\else
   \pdfoutput=1   
   \pdftrue
\fi

\def\la{\mathrel{\hbox{\rlap{\hbox{\lower4pt\hbox{$\sim$}}}\hbox{$<$}}}}
\def\ga{\mathrel{\hbox{\rlap{\hbox{\lower4pt\hbox{$\sim$}}}\hbox{$>$}}}}

\newcommand{\be}{\begin{eqnarray}}
\newcommand{\ee}{\end{eqnarray}}

\ifundefined{ensuremath}\def\ensuremath#1{\relax\ifmmode{#1}}
\else${#1}$\fi\else\relax\fi
\ifundefined{nuc}\def\nuc#1#2{\relax\ifmmode{}^{#1}{\protect\text{#2}}
\else${}^{#1}$#2\fi}\else\relax\fi

\newcommand{\kmps}{km~s$^{-1}$}

\newcommand{\msol}{\ensuremath{{\textrm{M}_\odot}}}

\newcommand{\nni}{\nuc{56}{Ni}}

\def\tstd{\ensuremath{\tau_{\textrm{std}}}}

\newcommand{\phx}{\texttt{PHOENIX}}
\newcommand{\snia}{SN~I\lowercase{a}}
\newcommand{\sneia}{SNe~I\lowercase{a}}
\newcommand{\gamray}{$\gamma$-ray}
\newcommand{\halpha}{H$\alpha$}
\newcommand{\hbeta}{H$\beta$}

\received{} 
\accepted{} 
\journalid{}{} 
\articleid{}{} 
\shortauthors{Lentz, E. et~al.}
\shorttitle{Detectibility of Hydrogen in Type I\lowercase{a} Supernovae}

\begin{document}

\title{Detectibility of Hydrogen Mixing in Type I\lowercase{a} Supernova
pre-Maximum Spectra}

\author{Eric J.~Lentz\altaffilmark{1}, E.~Baron\altaffilmark{2}, Peter
H.~Hauschildt\altaffilmark{1} and David Branch\altaffilmark{2}}

\altaffiltext{1}{Department of Physics and Astronomy \& Center for
Simulational Physics, University of Georgia, Athens, GA 30602, USA}
\email{lentz@physast.uga.edu, yeti@hal.physast.uga.edu}

\altaffiltext{2}{Department of Physics and Astronomy, University of
Oklahoma, 440 West Brooks, Norman, OK 73019-0260, USA}
\email{baron@nhn.ou.edu, branch@nhn.ou.edu}

\begin{abstract}
The presence of a small amount of hydrogen is expected in most single
degenerate scenarios for producing a Type Ia supernova
(SN~Ia). While hydrogen may be detected in very early high resolution
optical spectra, in early radio spectra, and in X-ray spectra, here we
examine the possibility of detecting hydrogen in early low resolution
spectra such as those that will be obtained by proposed large scale
searches for nearby SNe~Ia. We find that definitive detections will
require both very early spectra (less than 5 days after explosion) and
perhaps slightly higher amounts of hydrogen than are currently
predicted to be mixed into the outer layers of SNe~Ia. Thus, the
non-detection of hydrogen so far does not in and of itself rule out any
current progenitor models. Nevertheless,
very early spectra of SNe~Ia will provide significant clues to the
amount of hydrogen present and hence to the nature of the SN~Ia
progenitor system. Spectral coverage in both the optical and IR will
be required to definitively identify hydrogen in low resolution spectra.
\end{abstract}

\section{Introduction}

Type Ia supernovae (SNe~Ia) are now considered to be among the best
``standardizable candles'' available \citep{philetal99}, and they are
the tool of choice for observational cosmology
\citep{perletal99,riess_scoop98,riess00,goldhetal01}. However,
the progenitor system remains in doubt \citep[for a review
see][]{prog95}. The most widely accepted current view is that a C+O
white dwarf accretes hydrogen (or perhaps helium) from a companion
star until it reaches the Chandrasekhar mass and explodes due to the
thermonuclear burning of C+O. While the merging of two white dwarfs
(the double degenerate scenario) remains a viable progenitor for
SNe~Ia, recent theoretical work has focused on the accretion of
material via Roche lobe overflow from a MS or subgiant companion
(hydrogen cataclysmic variables) or via wind accretion from a
red-giant companion (symbiotic stars). A recent search for radio
emission appears to rule out the peculiar SN~Ia 1986G as having a
symbiotic star progenitor \citep{eck95}. \citet{maxted00} have found a
candidate double degenerate system containing a
0.5~\msol\ white dwarf and the companion is more massive that
0.97~\msol, so that the total mass of the system exceeds the
Chandrasekhar mass.  While it is clear that double degenerate
progenitors do exist, it is still not clear whether the merger of a
double degenerate system leads to collapse \citep[due to electron
capture,][]{sainom98,mochkoasi97,TY96} or explosion. It is also unclear
whether sufficiently massive double degenerate systems exist in
sufficient numbers to produce the observed sample of SNe~Ia
\citep{prog95}. Therefore the detection of hydrogen in a SN~Ia would
greatly expand our knowledge of the progenitor system. The question of
the identity of the progenitor system remains a significant hurdle for
the general acceptance of the observational cosmology results; is a
gap in our knowledge of the binary stellar evolution; and impedes our
understanding of the chemical evolution of galaxies. There have been
attempts to detect hydrogen via looking for narrow H$\alpha$ lines
\citep{cum94d96}, in the radio \citep{eck95}, and in X-rays
\citep{SP92A93,schlegelrpp95}.  While all of these methods should be
pursued, it seems likely that we will obtain large numbers of very
early low resolution optical spectra of nearby SNe~Ia from dedicated searches
that will begin taking data soon. Thus, we examine the possiblity of
detecting hydrogen in early SN~Ia spectra.

\section{Methodology}

The W7 model \citep{nomw7,nomw72} is a good match to normal \sneia\
spectra \citep{l94d01}, especially the outer layers visible during the
pre-maximum phase of the supernova. The outer layers of W7 are
unburned C+O enriched with solar metals, expanding at velocities,
$v > 15000$ \kmps, with a total mass of $\sim 0.07$~\msol.

In the single degenerate scenario, there are two sources of hydrogen:
1) circumstellar matter that is accreting onto the white dwarf from
the companion; 2) matter ablated from the companion by the supernova
explosion itself \citep*{ltw91,marietta00}. In the first case one
expects a relatively small amount of material that would most likely
be detected in high resolution spectra, radio, and in X-rays. In the
second case the amount of material is much larger and is at higher
velocity. It is more uncertain how it would be mixed into the outer
layers of the the supernova. It is the latter case that we examine
here in a parameterized way, guided by the numerical calculations of
\citet{ltw91} and \citet{marietta00}.

We have mixed solar composition material into the unburned C+O layers
at the top of the W7 explosion model.   We have replaced up to half of this
C+O mixture with solar composition material, while maintaining the
overall density structure and luminosity.  The models are homologously
expanded to account for the expansion since explosion to epochs of 5,
10, and 15 days after explosion, corresponding respectively to 15, 10,
and 5 days before maximum brightness using the canonical light curve
rise time of 20 days \citep[e.g.,][]{riessetalIaev00,akn00}.  We
calculate the \gamray\ deposition and decay of radioactive isotopes,
especially \nni, and use the luminosities for W7 models fit to
SN~1994D \citep{l94d01} at each epoch. The spectra and \gamray\
deposition are calculated using the general purpose radiative transfer
code \phx, version {\tt 11.9.0} \citep{hbjcam99}. \phx\ includes all of
the effects of special relativity in the steady-state solution of the
transport equation and energy balance.  We have included detailed NLTE
treatment for the ionic species: \ion{H}{1}, \ion{He}{1}--{II},
\ion{C}{1}--{IV}, \ion{O}{1}--{III}, \ion{Na}{1}--{III}, \ion{Mg}{1}--{IV},
\ion{Al}{1}--{IV}, \ion{Si}{1}--{V}, \ion{S}{1}--{IV}, \ion{Ca}{1}--{IV},
\ion{Fe}{1}--{VI}, and \ion{Co}{1}--{III}.

\section{Uniform Mixing in the Unburned Layer}

For the first set of numerical experiments, we have mixed solar
composition material throughout the unburned, outer, C+O layers of
W7. For each model, we have replaced 0.1, 1, 10, 30, or 50\% of the
C+O mixture mass with an equal mass of solar composition
material. Since the composition of the C+O material was designed to
have solar-like mass fractions for metals heavier than oxygen the mass
fraction of these heavier metals is not affected by the mixing
process. The 0.1\% and 1\% models had no detectable differences from
the unmixed models at any epoch.
 
\subsection{Day 15}

\begin{figure}
\begin{center}
\includegraphics[width=14cm,angle=0]{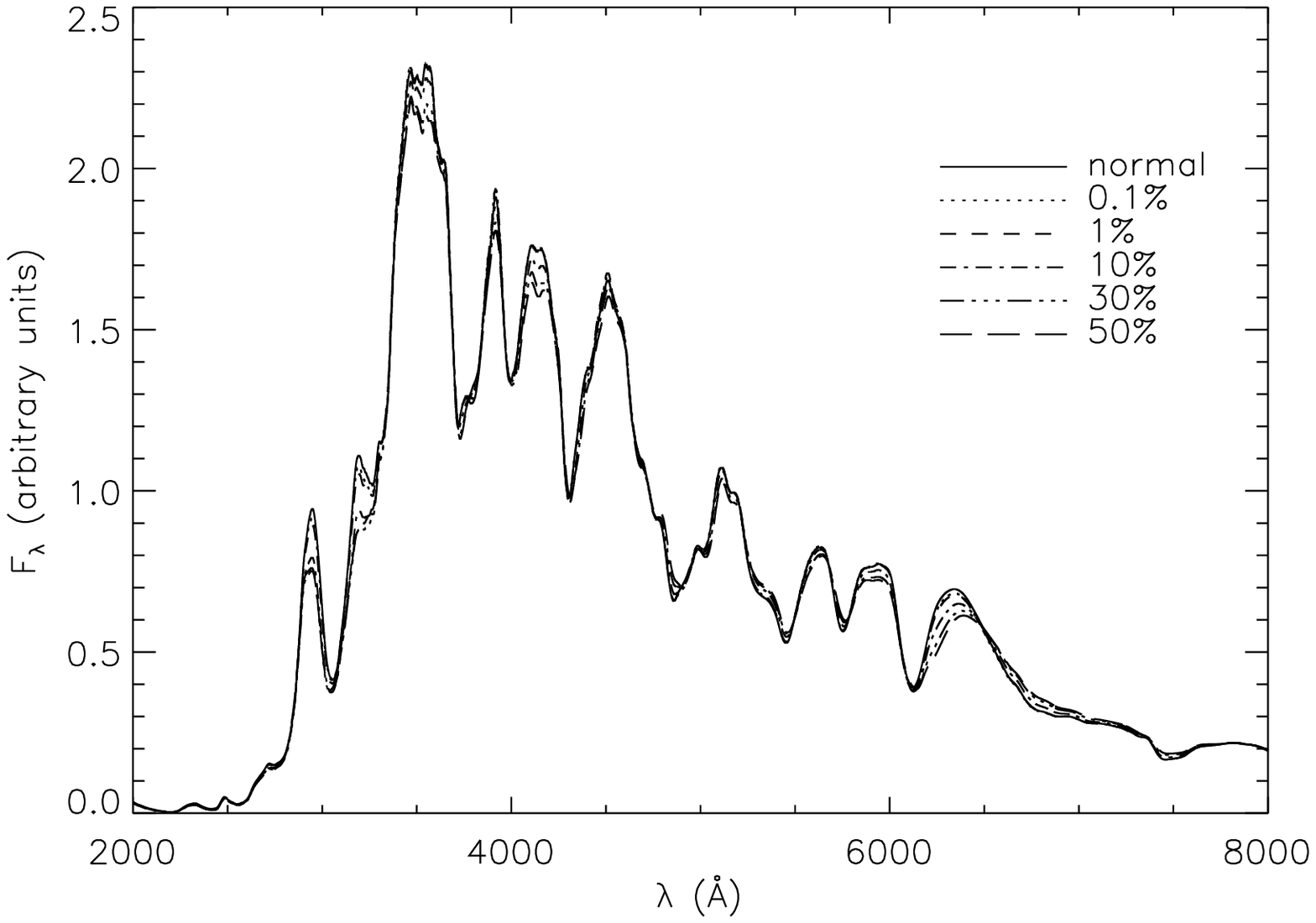}
\end{center}
\caption{The fully mixed models at Day 15. The legend labels the
percentage of hydrogen mixing which takes the values
(0,0.1,1,10,30,50)\% in this and the next two figures. 
\label{fig:day15full}}
\end{figure}

Our synthetic spectra for the mixing at day 15, or 5 days before
 maximum light, are shown in Figure~\ref{fig:day15full}. Other than
 some minor effects on the peaks in the blue part of the spectrum, the
 primary effect is the formation of a \halpha\ line at the red edge
 of the \ion{Si}{2}\ absorption at 6150~\AA. In the comparison of synthetic
 spectra, we can discern the effect for 10\% and larger mixing. Outside
 the controlled parameters of a numerical experiment this effect on
 the spectrum of the 10\% mixing model would likely not be recognized
 as \halpha. The 50\% mixing model does show a change in the shape of
 the 6150~\AA\ feature from what is normally seen, and would stand a
 reasonable chance of detection, while the 30\% model would be
 marginally detectable.
 For the detection of hydrogen, a single spectrum at 5 days before
 maximum will not be adequate.

\subsection{Day 10}

\begin{figure}
\begin{center}
\includegraphics[width=14cm]{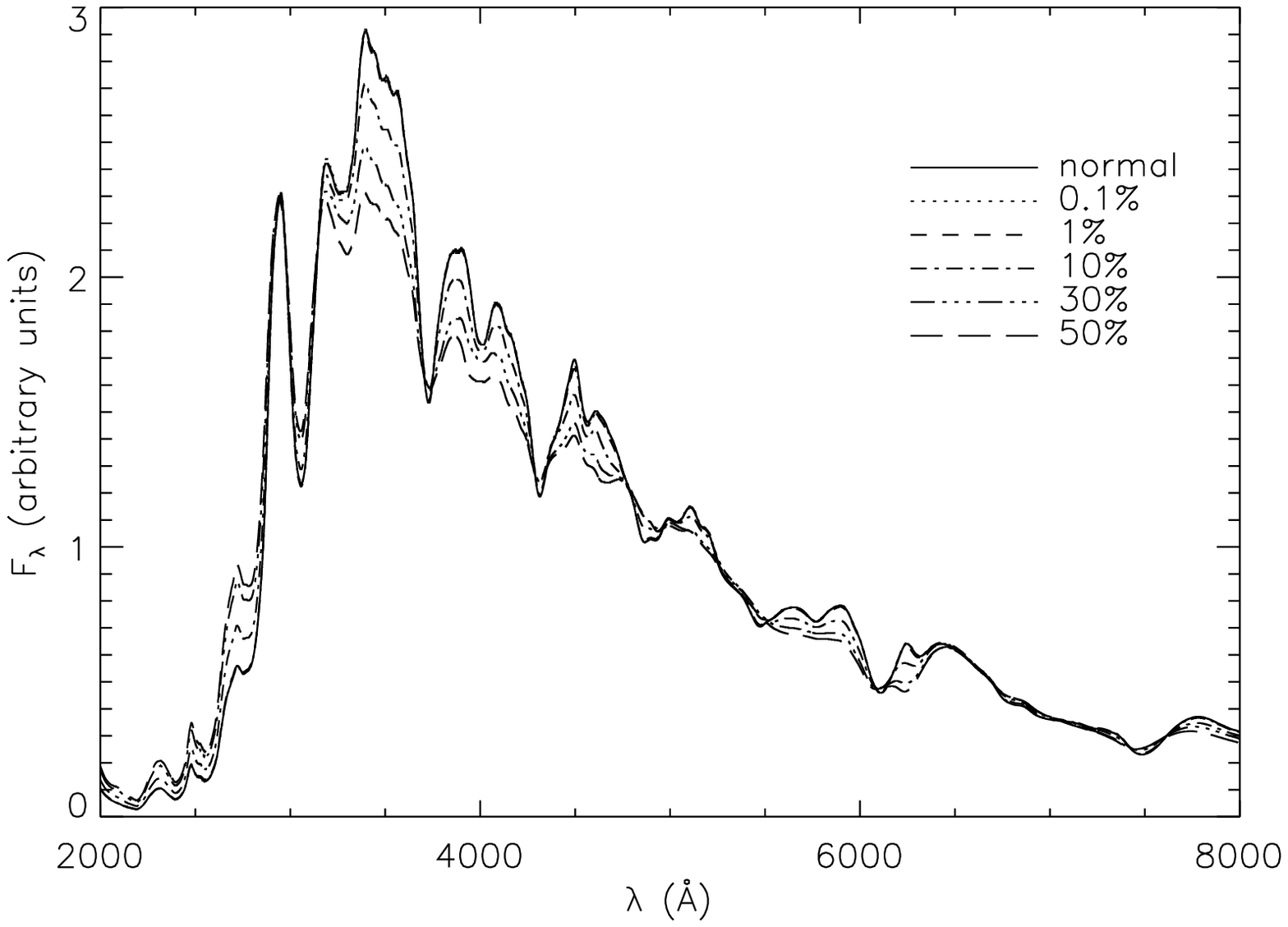}
\end{center}
\caption{The fully mixed models at Day 10. The legend labels the 
percentage of hydrogen mixing.  \label{fig:day10full}}
\end{figure}

At 10 days before maximum light, or 10 days after explosion
(Figure~\ref{fig:day10full}), the effects of mixing are
more pronounced and might be detected.  At 10 days after
explosion, the continuum optical depth at 5000~\AA, \tstd, at the base
of the C+O layer, is 0.15 for
the unmixed model, and is 0.5 for the 50\% mixed model. This increase
in electron scattering ``washes out'' the peaks and the troughs of
the spectral features.
\halpha\ produces a distinct double-bottom to the \ion{Si}{2} 6100~\AA\
feature. This is complicated by the presence of \ion{C}{2} absorption
at 6400~\AA. The \ion{C}{2} feature is seen faintly in some \sneia, at
these early epochs. Our previous models \citep{l94d01}, did not
include \ion{C}{2} in NLTE and did not show the feature. The 10\% mixed
model would likely be identified as \ion{C}{2} in an observed \snia. The
30\% and 50\% mixed models would be more likely to be identified as
\halpha\ if observed.  There is a modest absorption feature at
4600~\AA\ for \hbeta\ in the 30\% and 50\% mixed models that is buried
in the electron scattering changes in that region of the spectrum.

\subsection{Day 5}

\begin{figure}
\begin{center}
\includegraphics[width=14cm]{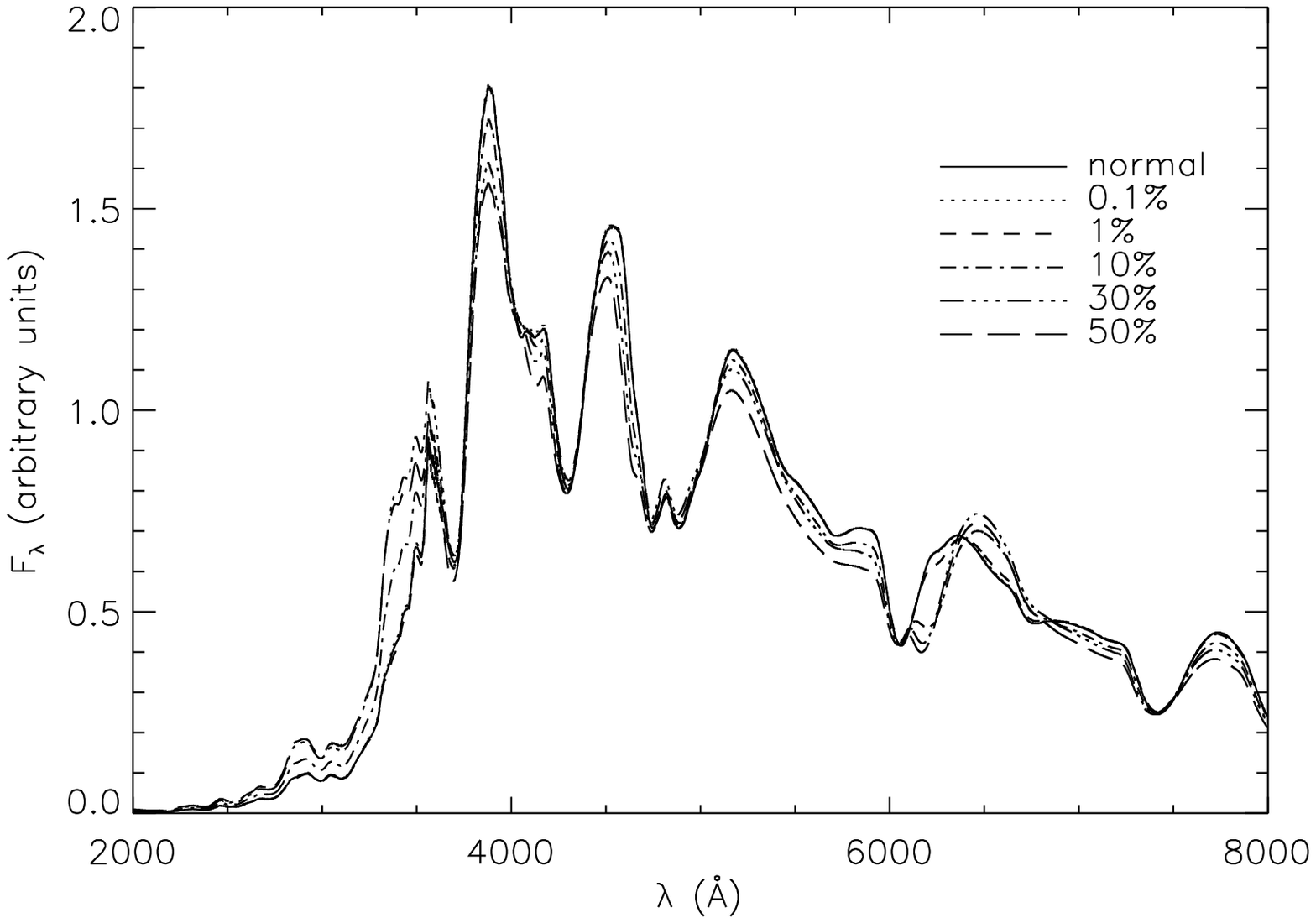}
\end{center}
\caption{The fully mixed models at Day 5. The legend labels the
percentage of hydrogen mixing.
\label{fig:day5full}}
\end{figure}

The earliest \sneia\ spectrum is for SN~1990N at
14 days before
maximum light \citep{leib91}.  We have calculated hydrogen mixed
models for W7 at day 5 (Figure~\ref{fig:day5full}),
which is nominally at 15 days before maximum light. Again, the effects
of the extra electron scattering from the additional electrons from
the hydrogen influences the spectrum. Because hydrogen has one
electron per proton, at a given mass density, the electron density is
higher in the solar composition mixture, than in the original C/O layer.
The effect of electron scattering is smaller at 5 days
after explosion than 10 
days for two reasons: the day 10 spectrum is bluer and the effects are
concentrated in the blue;  electron scattering is already strong
in the unmixed model at day 5, $\tstd = 0.6$ at 15000~\kmps. The
\halpha\ feature is quite strong and makes a definitive
double-bottomed absorption, with \ion{Si}{2} at 6200~\AA, for solar
mixing of 10\% and higher. The small indentation at the red peak of
the \ion{Si}{2} feature at 6000~\AA\ is due to \ion{C}{2}. The \halpha\
feature could be identified as a very strong \ion{C}{2} feature like
what might be expected in a C+O white dwarf merger scenerio, but one
would also expect to see other \ion{C}{2} features in that case. The
formation of this feature by carbon would require a significant mass
of carbon at high velocities, $v > 15000$~\kmps, in a white dwarf
merger model that produced otherwise normal \sneia\ spectra and light
curves. 

\begin{figure}
\begin{center}
\includegraphics[width=12cm,angle=90]{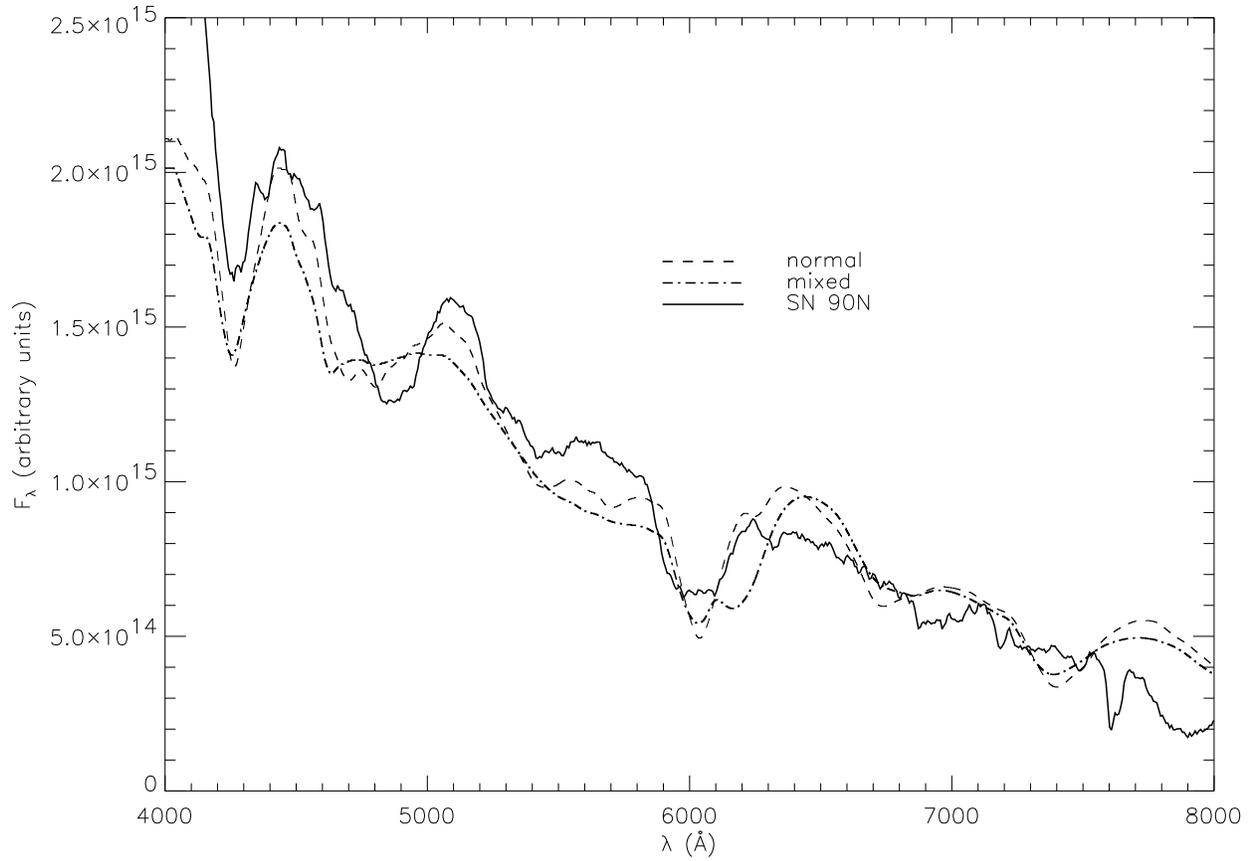}
\end{center}
\vspace{5pt}
\caption{The SN~1990N spectrum at day -14 \citep{leib91} is compared
to the spectra at day 5 (-15) with and without hydrogen
mixing. Clearly the hydrogen is too red to match the feature, although
the extra electrons do help somewhat to extend the blue wing of the silicon
line. The identification of high velocity carbon
\citep{fish90n96} is more probable than hydrogen.\label{fig:90nsi}}  
\end{figure}

Figure~\ref{fig:90nsi} compares the SN~1990N spectrum at day -14
\citep{leib91} to the spectra at day 5 (-15) with 50\% mixing of solar
composition material and without hydrogen
mixing. Clearly the hydrogen is too red to match the feature although
the extra electrons do help to extend the blue wing of the silicon
line. \ion{C}{2} creates an unseen bump in the middle of the
hydrogen/silicon absorption trough. The identification of silicon
mixed with 
carbon \citep{mazz90N01,fish90n96} is more probable than hydrogen.

\section{Variation of Mixing Depth}

We studied the depth of mixing of the solar material mixed into a
\snia\ since high velocity material would most likely come from the
companion and be mixed from the outside, but see \citet{marietta00}
for a detailed discussion of the expected hydrogen distribution. The
models in the previous section were all mixed to the depth of the
transition from partially burned to unburned C+O material at
15000~\kmps. 
In the full mixing models described above up to half
of this material is replaced with solar composition material. It is expected
\citep{marietta00} that the amount of mixing will vary with depth and
we address this effect here with a simple parameterization.

\begin{figure}
\begin{center}
\includegraphics[width=12cm,angle=90]{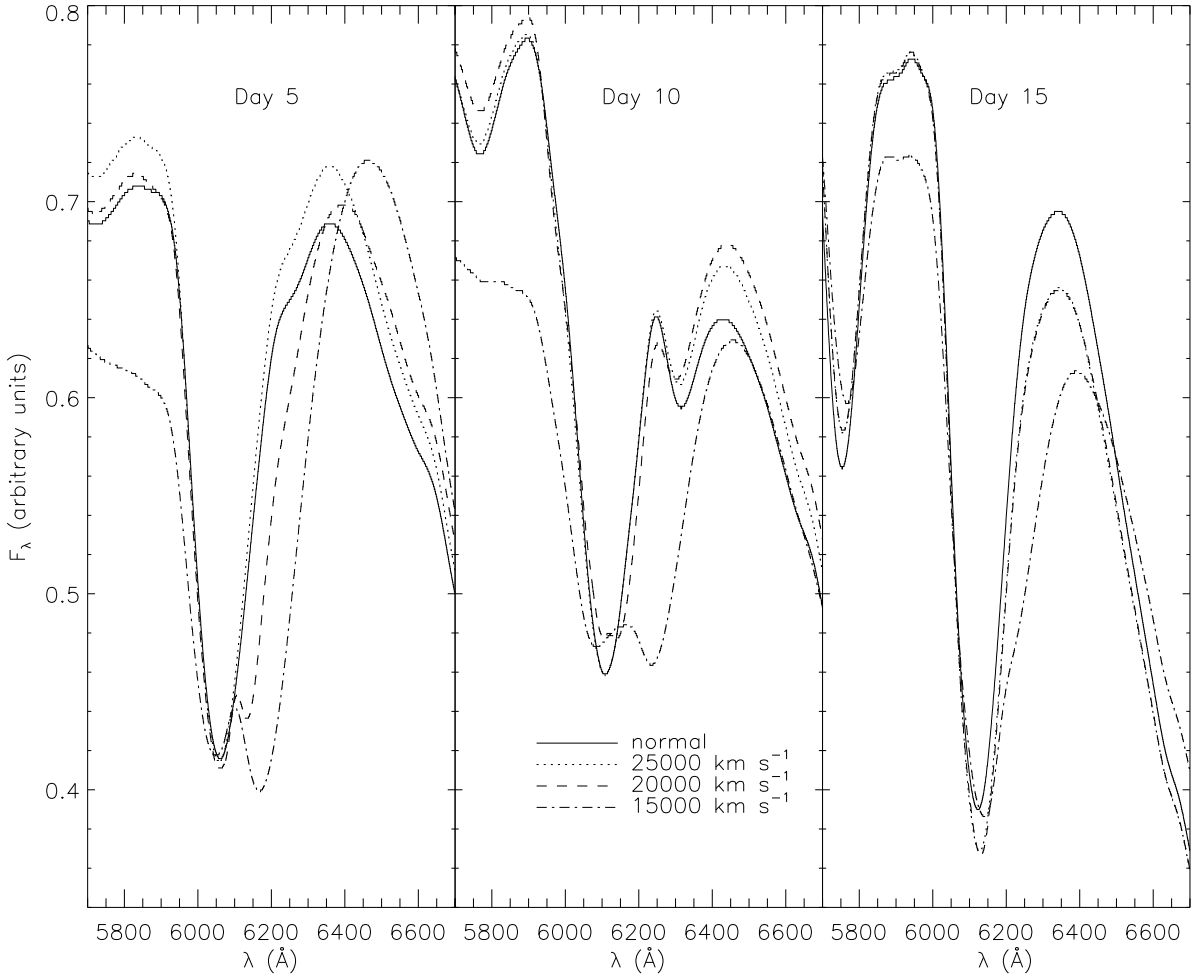}
\end{center}
\caption{The effects of mixing depth variation on the region around
the H$\alpha$ line at days
5, 10, and 15. Half of the C/O layer has been replaced with solar
composition material which has been mixed to depths of 
(15000, 20000, 25000)~\kmps\ as described in the legend.
\label{fig:mixdepth}}
\end{figure}

We have mixed the 50\% composition at days 5, 10, and 15, to depths of
25000~\kmps\ ($7.2 \times 10^{-4}$~\msol\ C+O above this velocity in
W7), 20000~\kmps\ ($7.2 \times 10^{-3}$~\msol), and 15000~\kmps\ ($7.2
\times 10^{-2}$~\msol, as stated above). Figure~\ref{fig:mixdepth}
shows the region of spectra around the H$\alpha$ feature for each of
these epochs. The mixing to only 25000~\kmps\ and 20000~\kmps\ does
not have detectable effects on the spectra with the minor exception of
the 20000~\kmps\ model at day 5 which is marginally detectable, and
shows evidence of a second feature affecting the \ion{Si}{2}.  An
observation of an event similar to this model is unlikely to be
conclusively recognized as evidence of hydrogen mixing in \sneia.
Only the models mixed substantially throughout the C+O layer show
clearly detectable modifications. \citet{thomas02} studied the shapes
of several \ion{Si}{2} lines near maximum light  (see their Figure 6),
and none of them showed features that would be indicative of the
presence of hydrogen.

\section{Conclusions}

The mixing of hydrogen and solar composition material into the outer
layers of a Type~Ia supernova model has little effect on the spectra,
unless the solar material replaces a substantial fraction of the
\snia\ C+O ejecta. In fact the non-detection of hydrogen in SNe~Ia is
expected in most of the present progenitor scenarios.  The effects of
mixing in the outer layers diminish as the photosphere recedes into
deeper layers of the ejecta. In our tests the most detectable results
were obtained by replacing half of the mass of the unburned C+O in the
outer layers with solar abundance material (for a total of $2.7 \times
10^{-2}$~\msol\ of hydrogen). These models produced reasonably strong
signals 5 and 10 days after the explosion, but are less prominent at
day 15, still 5 days before maximum light. One fifth of that mass of
mixed hydrogen is able to produce identifiable signals at 5 days after
explosion  and approximately half that mass is required 10 days
after the explosion.

The effects of additional electrons from the hydrogen atoms may also be
a useful diagnostic when coupled with detailed analysis of high
quality data. Replacing half of the C+O with solar material, but to
different depths, we found that only at 5 days after explosion did any
model other than mixing to the bottom of the C+O layer provide any
chance of detection. This is consistent with a picture based on the
mass of the mixed hydrogen.  Approximately 0.02~\msol\ of solar
composition material swept up from the wind of or ablated off of the companion
is needed for a detection at 10 days after explosion (10 days before
maximum light). The deeper it is mixed the later it will be
detectable.  For quantities of mixed hydrogen near the detection
threshold, the \ion{C}{2} feature in the red emission peak of
\ion{Si}{2} complicates the identification of \halpha\
absorption. Confirmation requires several epochs to trace the
development of the spectral features, and ideally another signal, such
as narrow circumstellar hydrogen lines. Even earlier spectra would be
advantageous.  

\begin{figure}
\begin{center}
\includegraphics[width=12cm,angle=90]{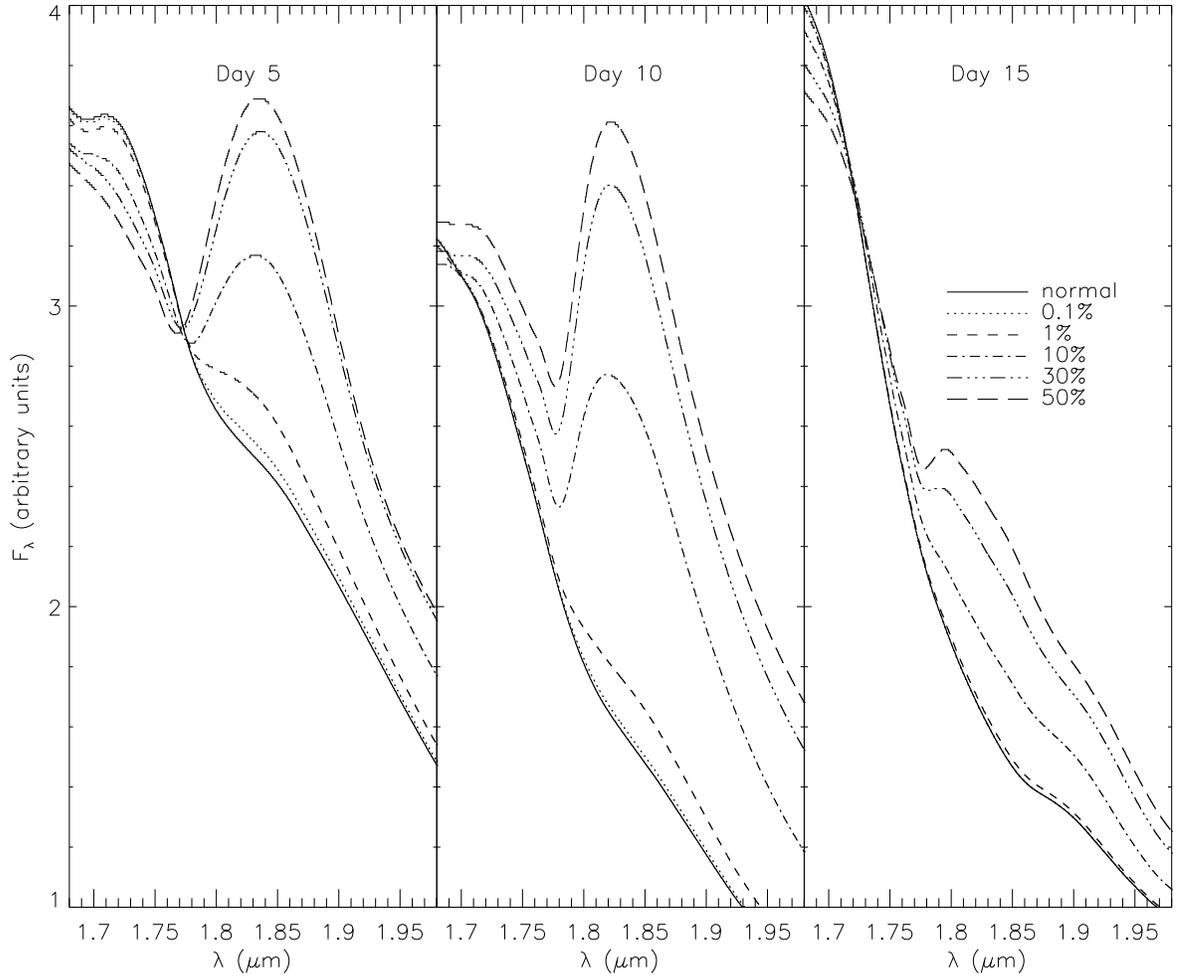}
\end{center}
\caption{The effects of mixing depth variation on the region around
the P$\alpha$ line at days
5, 10, and 15, for various mixing fractions (mixed to 15,000~\kmps) as
described in the legend. 
\label{fig:paschena}}
\end{figure}

Additional spectral signatures that would prevent confusion of
high-speed carbon ejecta that could mimic the \halpha\ from mixed
hydrogen are needed. Figure~\ref{fig:paschena} shows the region around
the P$\alpha$ line.  While the P$\alpha$ line shows a pretty strong
signature at the early epochs the P$\beta$ line is not easily
discerned, thus, combined IR and optical data would be likely be
required for a definitive detection. \citet{bowersetal97} and
\citet{hern98bu00} compared the IR data for SNe~Ia and most of the
data is obtained at late times (SN 1998bu has data at about 13 days
after maximum); the data are obtained much later than we desire and
without detailed spectral analysis it would be difficult to draw any
conclusions. SN~1999ee \citep{hamuy99ee99ex02} does have very early IR
spectra (approximately 11 days before maximum). It shows no sign of
any feature near P$\beta$, and the region around P$\alpha$ is not
covered spectroscopically. Nevertheless, early IR spectra are
obtainable and should be vigorously persued.

The results of \citet{marietta00} suggest that only about
$10^{-4}$~\msol\ of hydrogen would be stripped from the companion at
$v > 15,000$~\kmps, however, they find that $10^{-3}$~\msol\ of
hydrogen would be stripped from the companion at $v >
10,000$~\kmps. Thus, our results indicate that very early photospheric
spectra of SNe~Ia might be able to detect the presence of hydrogen,
and certainly could be used as a signal to activate high resolution,
radio, and X-ray programs.

Continued work in the radio, X-ray, and high resolution optical spectra can
verify and reinforce any putative hydrogen detections.

\acknowledgments This work was supported in part by NSF grant
AST-9720704, NASA ATP grant NAG 5-8425 and LTSA grant NAG 5-3619 to
the University of Georgia and by NASA grant
NAG5-3505, and an IBM SUR grant to the University of Oklahoma. PHH was
supported in part by the P\^ole Scientifique de Mod\'elisation
Num\'erique at ENS-Lyon. Some of the calculations presented here were
performed at the San Diego Supercomputer Center (SDSC), supported by
the NSF, and at the National Energy Research Supercomputer Center
(NERSC), supported by the U.S. DOE.  We thank both these institutions
for a generous allocation of computer time.


\clearpage

\end{document}